\documentclass{aastex631}
\usepackage{csquotes}
\usepackage{sidecap}
\sidecaptionvpos{figure}{t}
\begin{document}
\title{The JWST Resolved Stellar Populations Early Release Science Program. VIII. The Spatially Resolved Star Formation History of WLM}

\correspondingauthor{Roger E. Cohen}
\email{rc1273@physics.rutgers.edu}

\author[0000-0002-2970-7435]{Roger E. Cohen}
\affiliation{Department of Physics and Astronomy, Rutgers the State University of New Jersey, 136 Frelinghuysen Rd., Piscataway, NJ, 08854, USA}

\author[0000-0001-5538-2614]{Kristen B. W. McQuinn}
\affiliation{Department of Physics and Astronomy, Rutgers the State University of New Jersey, 136 Frelinghuysen Rd., Piscataway, NJ, 08854, USA}
\affiliation{Space Telescope Science Institute, 3700 San Martin Drive, Baltimore, MD 21218, USA}

\author[0000-0002-1445-4877]{Alessandro Savino}
\affiliation{Department of Astronomy, University of California, Berkeley, CA 94720, USA}

\author[0000-0002-8092-2077]{Max J. B. Newman}
\affiliation{Department of Physics and Astronomy, Rutgers the State University of New Jersey, 136 Frelinghuysen Rd., Piscataway, NJ, 08854, USA}

\author[0000-0002-6442-6030]{Daniel R. Weisz}
\affiliation{Department of Astronomy, University of California, Berkeley, CA 94720, USA}

\author[0000-0001-8416-4093]{Andrew E. Dolphin}
\affiliation{Raytheon, 1151 E. Hermans Rd.,
Tucson, AZ 85756}
\affiliation{Steward Observatory, University of Arizona, 933 N. Cherry Avenue, Tucson, AZ 85719, USA}

\author[0000-0003-4850-9589]{Martha L. Boyer}
\affiliation{Space Telescope Science Institute, 3700 San Martin Drive, Baltimore, MD 21218, USA}

\author[0000-0001-6464-3257]{Matteo Correnti}
\affiliation{INAF Osservatorio Astronomico di Roma, Via Frascati 33, 00078, Monteporzio Catone, Rome, Italy}
\affiliation{ASI-Space Science Data Center, Via del Politecnico, I-00133, Rome, Italy}

\author[0000-0002-7007-9725]{Marla C. Geha}
\affiliation{Department of Astronomy, Yale University, New Haven, CT 06520, USA}

\author[0000-0002-5581-2896]{Mario Gennaro}
\affiliation{Space Telescope Science Institute, 3700 San Martin Drive, Baltimore, MD 21218, USA}
\affiliation{The William H. Miller {\sc III} Department of Physics \& Astronomy, Bloomberg Center for Physics and Astronomy, Johns Hopkins University, 3400 N. Charles Street, Baltimore, MD 21218, USA}

\author[0000-0003-0394-8377]{Karoline M. Gilbert}
\affiliation{Space Telescope Science Institute, 3700 San Martin Drive, Baltimore, MD 21218, USA}
\affiliation{The William H. Miller {\sc III} Department of Physics \& Astronomy, Bloomberg Center for Physics and Astronomy, Johns Hopkins University, 3400 N. Charles Street, Baltimore, MD 21218, USA}

\author[0000-0002-3204-1742]{Nitya Kallivayalil}
\affiliation{Department of Astronomy, University of Virginia, 530 McCormick Road, Charlottesville, VA 22904, USA}

\author[0000-0003-1634-4644]{Jack T. Warfield}
\affiliation{Department of Astronomy, University of Virginia, 530 McCormick Road, Charlottesville, VA 22904, USA}

\author[0000-0002-7502-0597]{Benjamin F. Williams}
\affiliation{Department of Astronomy, University of Washington, Box 351580, U.W., Seattle, WA 98195-1580, USA}

\author[0000-0002-0372-3736]{Alyson M. Brooks}
\affiliation{Department of Physics and Astronomy, Rutgers the State University of New Jersey, 136 Frelinghuysen Rd., Piscataway, NJ, 08854, USA}

\author[0000-0003-0303-3855]{Andrew A. Cole}
\affiliation{School of Natural Sciences, University of Tasmania, Private Bag 37, Hobart, Tasmania 7001, Australia}

\author[0000-0003-0605-8732]{Evan D. Skillman}
\affiliation{Minnesota Institute for Astrophysics, University of Minnesota, Minneapolis, MN 55455, USA}

\author[0000-0001-9061-1697]{Christopher T. Garling}
\affiliation{Department of Astronomy, University of Virginia, 530 McCormick Road, Charlottesville, VA 22904, USA}

\author[0000-0001-9690-4159]{Jason S. Kalirai}
\affiliation{Johns Hopkins Applied Physics Laboratory, 11000 Johns Hopkins Road, Laurel, MD 20723, USA}

\author[0000-0003-2861-3995]{Jay Anderson}
\affiliation{Space Telescope Science Institute, 3700 San Martin Drive, Baltimore, MD 21218, USA}

\begin{abstract}
 Radial stellar population gradients within dwarf galaxies provide a promising avenue for disentangling the drivers of galaxy evolution, including environment.  Within the Local Volume, radial stellar age gradient slopes correlate with interaction history, contrary to model predictions, so dwarfs that are isolated provide a critical control sample.     
We measure radial stellar age gradients in the relatively isolated gas-rich dwarf irregular WLM, combining JWST NIRCam and NIRISS imaging with six archival Hubble fields over semi-major axis equivalent distances of 0$\lesssim$R$_{SMA}$$\lesssim$4 kpc ($\lesssim$3R$_{hl}$).  Fitting lifetime star formation histories (SFHs) to resolved color-magnitude diagrams (CMDs), radial age gradients are quantified using $\tau_{90}$ and $\tau_{50}$, the lookback times to form 90\% and 50\% of the cumulative stellar mass.  We find that globally, the outskirts of WLM are older on average, with ($\delta$$\tau_{90}$, $\delta$$\tau_{50}$)/$\delta$R$_{SMA}=$(0.82$^{+0.10}_{-0.10}$, 1.60$^{+0.23}_{-0.22}$) Gyr/kpc (stat.), in good agreement with simulations.  However, we also detect an azimuthal dependence of radial stellar age gradients, finding that stars on the leading edge of WLM (relative to its proper motion) are both younger and have a flatter age gradient compared to the trailing edge.  This difference persists over 0.6$\lesssim$R$_{SMA}$$\lesssim$3.2 kpc ($\sim$0.5$-$2.5R$_{hl}$) and lookback times up to $\sim$8 Gyr, and is robust to assumed stellar evolutionary model.  Our results are consistent with star formation triggered by ram pressure stripping from a circumgalactic and/or intergalactic medium, suggested by recent H\textsc{I}  
observations.  If confirmed, processes typifying dense environments, such as ram pressure stripping, may be more relevant to the evolution of isolated galaxies than previously thought.  
\end{abstract}

\section{Introduction \label{introsect}}

\subsection{Radial Stellar Age Gradients in Dwarf Galaxies \label{introgradsect}}

Dwarf galaxies, due to their low masses (M$_{\star}$$\lesssim$10$^{9}$M$_{\odot}$) and correspondingly shallow potential wells, are ideal laboratories to understand the drivers of galaxy evolution.    
Among these drivers, environment appears to play a prominent role, influencing several large-scale trends.  For example, position-morphology and morphology-density relationships are observed in the Local Volume such that dwarf irregular (dIrr) galaxies tend to be both more luminous and more isolated than dwarf spheroidals \citep[dSphs; e.g.,][]{einasto74,vandenbergh94a,vandenbergh94b,weisz11,weisz11b}.  
In addition, the fraction of dwarf galaxies with quenched star formation decreases farther from massive hosts \citep{geha12,weisz15,mao24}, and more isolated dwarfs tend to be more gas-rich \citep[e.g.,][]{mcconnachie12}.  

Observations of radial stellar population gradients \textit{within} dwarf galaxies further reinforce the idea that environment plays a critical role in their mass assembly histories.  In most dwarf galaxies, recent star formation is centrally concentrated, producing an \enquote{outside-in} radial stellar age gradient observed at the present day, in which the stellar populations in the outskirts are older on average than the central regions \citep[e.g.,][]{sarajedini97,dohmpalmer98,gallagher98,lee99,sl99,aparicio00,aparicio00b,drozdovsky00,hidalgo03,cannon03,alonsogarcia06,bernard07,dejong08,hidalgo08,hidalgo13,delpino13,santana16,mcquinn17,bettinelli19,savino19}.  However, there are exceptions to this rule.  Such exceptions include NGC 6822, which has a flat radial stellar age gradient \citep{cannon12,fusco14}, as well as M33 and the Large Magellanic Cloud at the more massive end, with a \enquote{V-shaped} radial age trend characterized by an inversion where the age gradient changes from inside-out to outside-in \citep{williams09,cohen24}.  Strikingly, all of these exceptions have evidence for past interactions \citep[][but see \citealt{bernard12}]{zhang21,choi22,smercina23}, underscoring the role of environment in dwarf galaxy mass assembly.

The idea that environment is a driving factor in setting present-day stellar age gradients is at odds with predictions from the latest simulations.  Two independent sets of cosmological hydrodynamical zoom-in simulations by \citet{graus19} and \citet{riggs24} examined age gradients in dwarf galaxies.  While only \citet{riggs24} examined environment, both find that stellar age gradient slopes correlate with the lifetime assembly history of the galaxy as a whole.  The simulations predict that globally older galaxies, having formed the majority of their mass at earlier lookback times, display steeper outside-in present-day age gradients.  On the other hand, globally younger galaxies, with larger fractions of their cumulative mass formed recently, should show flatter radial stellar age gradients.  In both sets of simulations, the lack of a relationship between stellar age gradients and environment persists because present-day age gradients are primarily a consequence of internal processes driven by stellar feedback (see \citealt{collinsread} for a review).  In contrast to earlier N-body smoothed particle hydrodynamics simulations \citep{stinson09,schroyen13}, the simulated dwarfs actually form inside-out, similar to their more massive counterparts \citep[e.g.,][]{matteucci89,bird13}.  Over time, feedback-driven processes then reshuffle stellar orbits radially in the \citet{graus19} and \citet{riggs24} simulations, preferentially driving older stars to more external radii to yield present-day outside-in age gradients (also see earlier results by \citealt{stinson09,elbadry16}).  For globally younger galaxies, the simulations predict that late-time star formation can proceed at large galactocentric radii, 
counteracting the radial reshuffling and flattening age gradients.  

Dwarfs that are isolated serve as critical testbeds for model predictions of radial stellar population gradients, 
and the impact of environment on galaxy evolution in general. By measuring radial stellar age gradients in relatively isolated galaxies where the impact of environment and galaxy-galaxy interactions is minimal, we may isolate the role of internal processes in setting present-day stellar age trends.
However, there are only a handful of cases where radial age gradients have been measured quantitatively rather than qualitatively to allow a comparison to simulations.  \citet{hidalgo13} found age gradient slopes ranging from outside-in to flat in four relatively isolated Local Group dwarfs (Cetus, Tucana, Phoenix, LGS-3), in addition to outside-in-gradients in mean age detected in Fornax, Sculptor and Leo I \citep{delpino13,bettinelli19,ruizlara21}.  Furthermore, these studies were based on low-mass, gas-poor dwarf spheroidals and transition dwarfs and lacked currently star-forming dIrr galaxies \citep[e.g.][]{mcconnachie12}.  Therefore, observations of more massive, isolated gas-rich dIrrs provide critical tests of both simulation predictions and observed trends.

\subsection{WLM as a Test Case \label{wlmtestsect}}

Within the Local Group, WLM (Wolf-Lundmark-Melotte = DDO 221) serves as the archetypal template of an isolated, gas-rich dIrr, and its properties are summarized in Table \ref{proptab}.  
At a distance of D$_{\odot}$=968$^{+41}_{-40}$ kpc \citep[yielding a scale of 282 pc/arcmin;][]{albers19}, WLM currently lies $>$800 kpc from either the Milky Way or M31, at $\gtrsim$3 times their virial radii.\footnote{We conservatively assume virial radii of R$_{\rm vir}$$\approx$300 kpc for both the Milky Way and M31, with recent estimates providing slightly ($\sim$10$-$20\%) lower values \citep[e.g.,][]{fardal13,watkins19,putman21}.}  Furthermore, WLM is likely on its first passage through the Local Group, and was even more isolated in the past based on recent proper motion and radial velocity measurements (\citealt{bennet24}, also see \citealt{battaglia22}).  Previous imaging studies of WLM have noted its composite stellar population, with a broad range of ages, as well as its outside-in stellar radial age gradient \citep{ferraro89,minniti97,rejkuba00,dolphin_wlm}, but this gradient was never quantified.  A direct comparison of an inner and outer field in WLM was made by \citet{albers19} based on Hubble Space Telescope (HST) observations.  Imaging of both the inner field, obtained using the Advanced Camera for Surveys Wide Field Channel (ACS/WFC) and the outer field, obtained with the Ultraviolet/Visible channel of Wide Field Camera 3 (WFC3/UVIS), extended faintward to reach the ancient ($\sim$13 Gyr) main sequence turnoff, allowing a detailed comparison of the two fields.  By fitting lifetime star formation histories (SFHs) to resolved color-magnitude diagrams (CMDs) in each field (see \citealt{tolstoy09} and \citealt{annibalitosi} for reviews), \citet{albers19} were able to demonstrate that the more external UVIS field is substantially older than the more internal ACS field.  However, acknowledging the limited spatial coverage of then-available observations, they refrained from quantifying radial age gradients in WLM. 

Fortunately, the available coverage of WLM using high-fidelity space-based imagers has since improved.  
Critical to our study, WLM was targeted by the JWST Resolved Stellar Populations Early Release Science (ERS) program (PID: ERS-1334) due in part to a variety of extant archival HST imaging.  An overview of our ERS program is given in \citet{ers_overview}, and in brief, has imaged three targets (M92, WLM, and the Draco dSph) using a single NIRCam pointing and a single parallel NIRISS pointing per target.  Using these observations, the ERS program has thus far provided the community with software for precision photometry of JWST images \citep{ers_dolphot} as well as detailed recommendations for the use of the JWST exposure time calculator \citep{savino24}, and photometric selection of C- and M-type asymptotic giant branch stars \citep{boyer24} and against background galaxy contaminants \citep{warfield23}.  In addition to the two deep fields analyzed by \citet{albers19}, our JWST NIRCam and NIRISS pointings are complemented by several additional fields imaged by Hubble (described in Sect.~\ref{obssect}), together providing the requisite spatial coverage to measure radial stellar population gradients.  \citet{ers_kristy} exploited the spatial overlap between the ACS/WFC field studied by \citet{albers19} and the JWST NIRCam field to perform a direct comparison between the resulting SFHs, finding excellent agreement.  Here, we leverage the ability to combine SFH results from the flagship imagers onboard both HST and JWST to examine spatial variations in the SFH across WLM as a function of both distance and azimuthal position angle relative to its center.

An additional advantage provided by WLM for CMD-based SFH studies is the presence of independent empirical constraints on its chemical evolution history at both early and late times.  \citet{leaman13} report $\langle$[Fe/H]$\rangle$=$-$1.46 (on the \citealt{carretta09} metallicity scale) from spectroscopy of red giants, with a standard deviation of $\sigma$=0.38$\pm$0.04 dex.  An estimate of the stellar metallicity distribution at ancient times is also provided by \citet{sarajedini23} based on the periods of ab-type RR Lyrae variables in the inner \citet{albers19} field, finding an asymmetric metallicity distribution with $\langle$[Fe/H]$\rangle$=$-$1.74$\pm$0.02 and a best-fit Gaussian peak at [Fe/H]=$-$1.60 with $\sigma=0.14$ dex (on the same metallicity scale).  Present-day abundances from HII regions are 12+Log$_{\rm 10}$ (O/H)=7.83$\pm$0.06 ([O/H]=$-$0.83; \citealt{lee05}), in good agreement with earlier measurements \citep{hm95,skillman89} as well as spectroscopy of O-B-type supergiants \citep{bresolin06}.  Stellar metallicities of $-$1.0$\lesssim$[M/H]$\lesssim$$-$0.5 with a mean of [M/H]=$-$0.87$\pm$0.06 were obtained from non-LTE analyses of six B-A-type supergiants \citep[][also see \citealt{venn03}]{urbaneja08}, implying solar [$\alpha$/Fe] ratios.  These metallicity constraints at both early and late times are in good agreement with the age-metallicity relation independently recovered from SFH fits to photometry of the NIRCam field by \citet{ers_kristy}, yielding a best-fit SFH in excellent agreement with our results based on updated photometry (see Sect.~\ref{modelcompsect}).     

Here, we quantify radial stellar age gradients in WLM, leveraging the depth and coverage of available HST and JWST imaging with available metallicity constraints to measure spatially resolved SFH trends directly from resolved CMDs.  We provide the first quantitative measurement of radial stellar age gradients \textit{in an isolated dIrr}, facilitating the first comparison to as-yet-untested simulation predictions.   
This study is organized as follows: In Sect.~\ref{datasect}, we describe our HST and JWST imaging fields, our technique for precision photometry, and present CMDs.  In Sect.~\ref{anasect} we present our method for measuring radial stellar age gradients from spatially resolved SFH fits to CMDs.  In Sect.~\ref{resultsect}, we quantify radial age gradients in WLM at different orientations and discuss potential causes, and in Sect.~\ref{summarysect} we summarize our results and directions for further investigation.  

\begin{deluxetable}{lrc}
\tablecaption{Summary of WLM Properties \label{proptab}}
\tablehead{
\colhead{Parameter} & \colhead{Value} & \colhead{Reference}}
\startdata
Morphological type & dIrr & (1) \\
RA$_{\rm J2000}$ (hh:mm:ss.ss) & 00:01:58.53 & (2) \\
Dec$_{\rm J2000}$ (-dd:mm:ss.ss) & -15:27:27.20 & (2) \\
M$_{V}$ (mag) & -14.2 & (1) \\
$E(B-V)$ (mag) & 0.03 & (3) \\
$(m-M)_{0}$ (mag) & 24.93$\pm$0.09 & (4) \\
D$_{\odot}$ (kpc) & 968$^{+41}_{-40}$ & (4) \\
D$_{\rm MW}$ (kpc) & 967$\pm$41 & (4) \\
D$_{\rm M31}$ (kpc) & 857$\pm$29 & (5) \\
$\langle$[M/H]$\rangle$ of supergiants (dex) & $-$0.87$\pm$0.06 & (6) \\
$\langle$[Fe/H]$\rangle$ of red giants (dex) & $-$1.46; $\sigma$=0.38$\pm$0.04 & (7) \\
$\langle$[Fe/H]$\rangle$ of RR Lyrae variables (dex) & $-$1.74$\pm$0.02 & (8) \\
Total stellar mass (M$_{\star}$/M$_{\odot}$) & 4.3$\times$10$^{7}$ & (9) \\
  & 2.6$\times$10$^{7}$ & (10) \\
Total H\textsc{I} mass (M$_{\rm HI}$/M$_{\odot}$) & 6.3$\pm$0.3$\times$10$^{7}$ & (9) \\
 & 5.5$\pm$0.3$\times$10$^{7}$ & (11) \\
Position angle (deg E of N) & 183.58 & (2) \\
Axis ratio $b/a$ & 0.44 & (2) \\
Half-light radius R$_{hl}$ ($\arcsec$) & 277.8 & (2) \\
Exponential disk scale length $h_{r}$ ($\arcsec$) & 131.7 & (2) \\
\enddata
\tablecomments{References: (1) \citet{mcconnachie12}; (2) McQuinn et al.~(2024c, in prep.); (3) \citet{sf11}; (4) \citet{albers19}; (5) \citet{degrijs14}; (6) \citet{urbaneja08}; (7) \citet{leaman13}; (8) \citet{sarajedini23}; (9) \citet{kepley07}; (10) \citet{cook14a}; (11) \citet{ian20}}
\end{deluxetable}

\section{Data \label{datasect}}

\subsection{Observations \label{obssect}}

\begin{figure}
\gridline{\fig{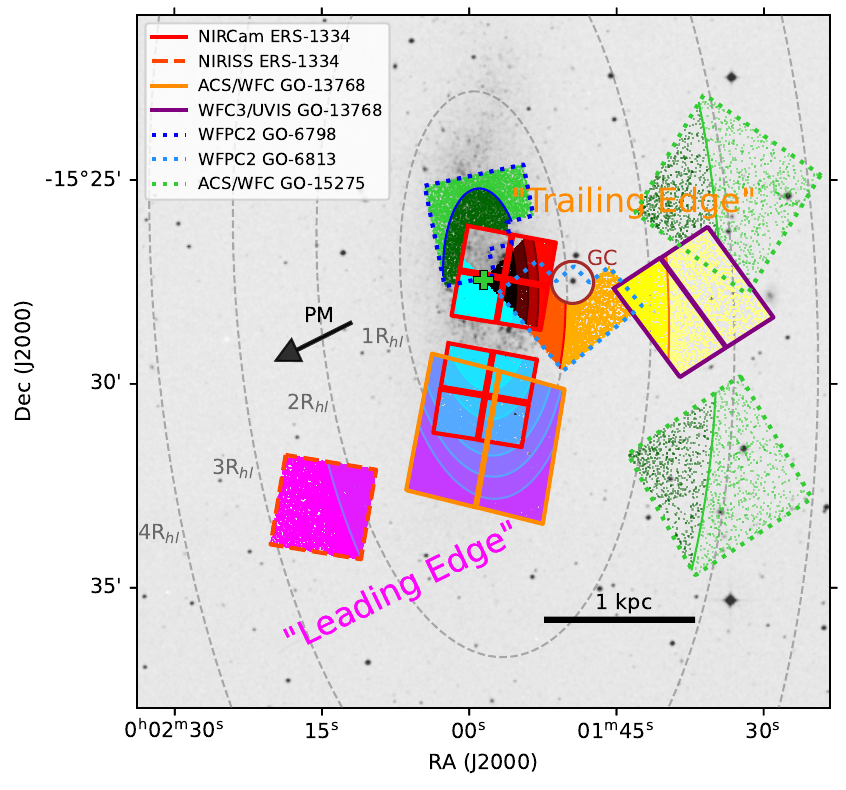}{0.85\textwidth}{}}
\caption{Footprints of the HST and JWST pointings analyzed here, overplotted on a DSS2-blue image of WLM 17$\arcmin$ per side.  For clarity, dotted lines indicate the footprints of fields with shallower photometry (see Sect.~\ref{photsect}).  Within each footprint, thin curved lines and colored points indicate the individual elliptical annuli in which SFHs were fit independently to measure lifetime SFH trends in WLM as a function of radius and position angle relative to its center (shown as a green cross).  The subsets of these annuli included in the \enquote{leading edge} and \enquote{trailing edge} samples analyzed in Sect.~\ref{radgradsect} are shown using magenta-blue and black-yellow color-coding respectively for their individual observed sources.  For the shallow off-axis fields excluded from the leading and trailing edge samples, light and dark green are used to indicate sources in each of the two elliptical annuli per pointing.  The arrow indicates the direction of WLM's proper motion from \citet{bennet24}.  The brown circle indicates the tidal radius of the single known globular cluster in WLM, within which sources were excluded from analysis.  Dotted grey lines indicate semi-major axis equivalent distance from the center of WLM in terms of the exponential half-light radius R$_{hl}$$=$278$\arcsec$ measured from ellipse fits to cleaned 3.6$\mu$ imaging.  
The scale bar in the lower right corresponds to 1 kpc for an assumed WLM distance modulus of $(m-M)_{0}=$24.93 mag \citep{albers19}.} 
\label{obs_fig}
\end{figure}

One factor driving the selection of WLM as a target for the JWST ERS stellar populations program was the substantial amount of HST imaging already available.  This archival imaging of WLM consists of a variety of programs using various filter/instrument combinations and photometric depths (e.g., Fig.~1 of \citealt{ers_dolphot}).  
Our goal here is to characterize spatial variations in the lifetime SFH of WLM, so we include in our analysis only the datasets amenable to SFH fitting, with imaging in multiple optical or near-infrared broadband filters yielding CMDs extending faintward of the red clump.  Six archival HST fields met these criteria, and their placement, along with our JWST NIRCam and NIRISS pointings, is shown in Fig.~\ref{obs_fig}.  Basic observational information for each field, including filters and total exposure times, is given in Table \ref{obstab}.  

\begin{deluxetable}{lcccccc}
\tablecaption{Observed Fields \label{obstab}}
\tablehead{
\colhead{Observatory} & \colhead{Camera} & \colhead{PID} & \colhead{RA (J2000)} & \colhead{Dec. (J2000)} & \colhead{Filters} & \colhead{t$_{exp}$} \\ \colhead{} & \colhead{} & \colhead{} & \colhead{$^{\circ}$} & \colhead{$^{\circ}$} & \colhead{} & \colhead{sec}}
\startdata
JWST & NIRCam & ERS-1334 & 0.489289 & -15.4812472 & F090W, F150W & 30492, 23707 \\
JWST & NIRISS & ERS-1334 & 0.561902 & -15.5504041 & F090W, F150W & 26670, 19842 \\
HST  & ACS/WFC & GO-13768 & 0.488824 & -15.5228656 & F475W, F814W & 27360, 34050 \\
HST  & WFC3/UVIS & GO-13768 & 0.407325	& -15.4684236 & F475W, F814W & 27360, 34050 \\
HST & WFPC2 & GO-6813 & 0.457812 & -15.4669512 & F555W, F814W & 5300, 5400 \\
HST & WFPC2 & GO-6798 & 0.494764 & -15.4340407 & F555W, F814W & 4900, 4900 \\
HST & ACS/WFC & GO-15275 & 0.393522 & -15.4201397 & F606W, F814W & 2325, 2187 \\
HST & ACS/WFC & GO-15275 & 0.395670 & -15.5344555 & F606W, F814W & 2324, 2185 \\
\enddata
\end{deluxetable}

We characterized the distance of each field from the center of WLM using the semi-major axis equivalent distance R$_{SMA}$. To calculate R$_{SMA}$, we assumed a WLM center location of (RA, Dec)$_{\rm J2000}$$=$(00:01:58.53, -15:27:27.20), an axis ratio of $b/a=$0.44 and a position angle of 183.58$^{\circ}$ east of north.  These values are based on ellipse fits to 3.6$\mu$ \textit{Spitzer} imaging meticulously cleaned of foreground contaminants (described in McQuinn et al.~2025, in prep.), yielding an exponential scalelength $h_{r}$=131.7$\arcsec$ (0.68 kpc) and half-light radius R$_{hl}$=277.8$\arcsec$ (1.30 kpc), consistent with other estimates \citep{rc3,hunter06,mcconnachie12,leaman12,cook14b,higgs21}.       

\subsection{Photometry \label{photsect}}

We apply the same analysis tools and techniques to generate photometric catalogs for both HST and JWST imaging.  We use the publicly-available \texttt{DOLPHOT} software package \citep{dolphot,dolphot2,ers_dolphot} to perform additional preprocessing and point-spread function fitting (PSF) photometry for each field independently.  \texttt{DOLPHOT} includes camera- and filter-specific model PSFs for imagers onboard HST and JWST, and produces photometric catalogs in the Vegamag photometric system.  To prepare individual science images\footnote{For ACS/WFC and WFC3/UVIS pointings, individual science images are \texttt{flc} format (corrected for charge transfer inefficiency), and \texttt{cal} images for NIRCam and NIRISS.  For WFPC2, data quality arrays are provided separately in \texttt{c1m} format, and applied to \texttt{c0m} science images.} for PSF photometry, several additional preprocessing steps are performed within \texttt{DOLPHOT} to mask bad pixels, apply pixel area masks, split images into individual chips and generate initial sky background frames.  A deep, distortion-corrected drizzled reference image is required to cross-match sources across science images, which we generate for HST pointings using the \texttt{drizzlepac} python package \citep{drizzlepac1,drizzlepac2}, while for NIRCam and NIRISS the Level 3 calibrated \texttt{i2d} images produced by the JWST pipeline are used as reference images.  There are various parameters within \texttt{DOLPHOT} that affect the details of how PSF photometry is performed, and we use the parameters recommended by \citet{williams_phat} for HST and \citet{ers_dolphot} for JWST, noting that the JWST observations we analyze served as a testbed for optimizing these parameters across various crowding conditions in the latter study.  

The photometric catalogs produced by \texttt{DOLPHOT} contain several diagnostic parameters used to cull poorly detected and non-stellar sources, described in the \texttt{DOLPHOT} manual\footnote{\url{http://americano.dolphinsim.com/dolphot/}}.  For HST imaging, we retain only sources with object type $\leq$2 and per-filter signal-to-noise ratio SNR$>$4, photometric quality flag $<$4, (sharp$_{\rm 1}$+sharp$_{\rm 2}$)$^{2}$$\leq$0.075 and (crowd$_{\rm 1}$+crowd$_{\rm 2}$)$\leq$0.1, where subscripts denote the per-filter values in each of the two filters.  For NIRCam and NIRISS imaging, our photometric quality cuts follow the recommendations of \citet{ers_dolphot}, requiring object type $\leq$2 and per-filter SNR$\geq$4, sharp$^{2}$$\leq$0.01, crowd$\leq$0.5 and photometric quality flag $<$4.  Because our SFH fitting procedure requires a model of observational noise and incompleteness, we perform artificial star tests using \texttt{DOLPHOT}, inserting $>$10$^{6}$ artificial stars per field.  Artificial stars are inserted and recovered individually, so there is no practical restriction on their proximity to each other, and each artificial star is considered recovered if it passes all of the photometric quality cuts described above.  

\subsubsection{JWST NIRCam and NIRISS Photometry}

\begin{figure}
\gridline{\fig{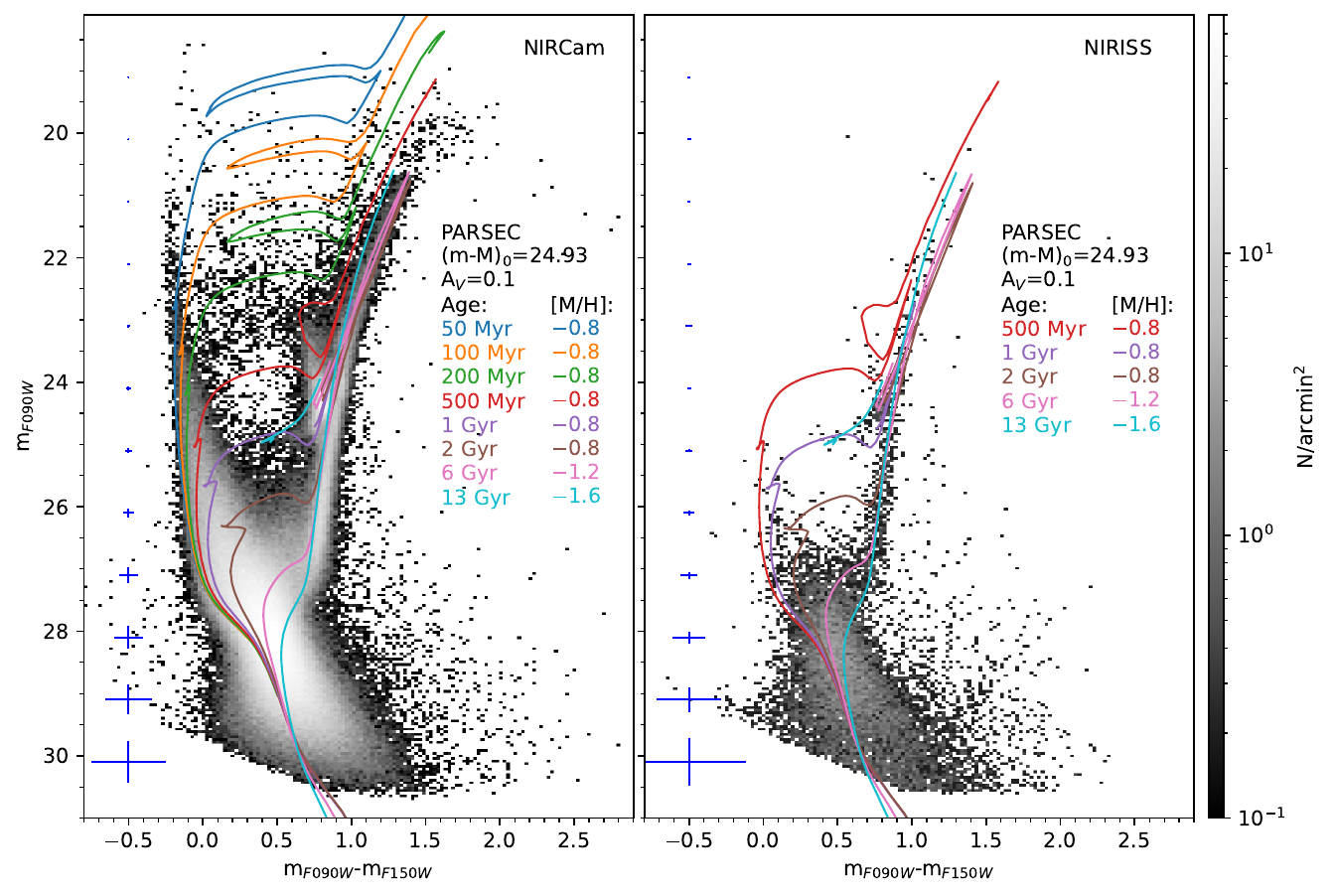}{0.97\textwidth}{}}
\vspace{-0.8cm}
\gridline{\fig{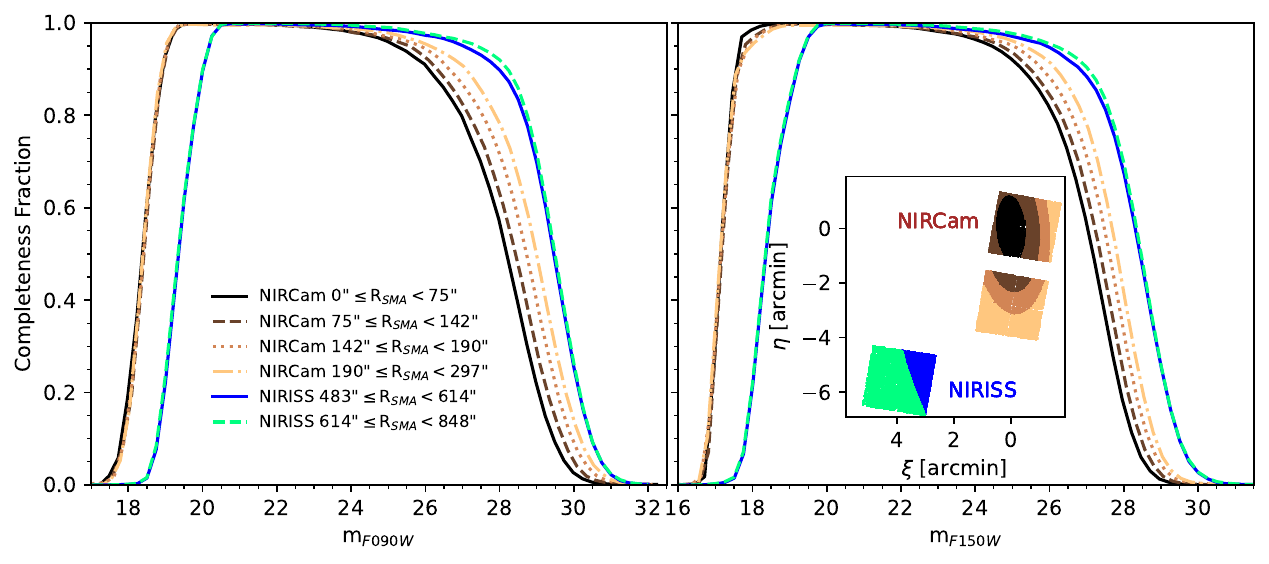}{0.8\textwidth}{}}
\vspace{-0.75cm}
\caption{\textbf{Top row:} CMDs of the WLM NIRCam (left) and NIRISS (right) fields, greyscaled by projected stellar density as indicated by the colorbar along the right-hand side.  Median photometric errors in magnitude bins are shown in blue along the left-hand side of each panel, and representative PARSEC isochrones \citep{parsec} are overplotted (see Sect.~\ref{wlmtestsect}), using the same color-coding in both panels.  The observed CMD loci of the stellar evolutionary sequences are in good agreement with the models for both NIRCam and NIRISS F090W and F150W imaging, and the predominance of old stellar populations in the NIRISS field is immediately apparent. \textbf{Bottom row:} Photometric completeness versus magnitude in the F090W (left) and F150W (right) filters of NIRCam and NIRISS, assessed using the artificial star tests.  Completeness curves are shown in equally-populated radial bins for each instrument, with semi-major axis equivalent distances given in the left-hand panel, and the radial bin locations illustrated in the inset of the right-hand panel.  For the more centrally-located NIRCam field, the impact of crowding across the detector is clear, while in the more externally located NIRISS field, lower projected stellar density yields similar photometric completeness regardless of location.} 
\label{jwstcmd_fig}
\end{figure}

In the top row of Fig.~\ref{jwstcmd_fig}, we compare CMDs from the NIRCam field, which covers a central region of WLM, to the NIRISS field located more externally, $\sim$5.2$\arcmin$ ($\sim$1.5 kpc) to the southeast.  The NIRISS field shows a clear lack of young stars present in the NIRCam field.  The apparent central concentration of recent star formation is qualitatively similar to the results of \citet{albers19}, who found a similar lack of young stellar populations in a UVIS field located $\sim$4.9$\arcmin$ ($\sim$1.4 kpc) to the west compared to a more internally located ACS field.  However, the two external (UVIS and NIRISS) fields are located at drastically different position angles ($\sim$145$^{\circ}$ and $\sim$264$^{\circ}$ east of north respectively; see Fig.~\ref{obs_fig}), and we discuss their star formation histories quantitatively in Sect.~\ref{resultsect}.

In the NIRCam and NIRISS CMDs in the top row of Fig.~\ref{jwstcmd_fig}, we have overplotted PARSEC isochrones \citep{parsec} for a range of ages and metallicities appropriate to WLM \textbf{(e.g., \citealt{urbaneja08,leaman13,sarajedini23}, see Sect.~\ref{wlmtestsect})}.  Thanks to the depth and dynamic range of our JWST imaging, several features are apparent.  First, the model-predicted loci of intermediate to old ($\gtrsim$Gyrs) stellar populations are qualitatively in good agreement with observations for both the NIRCam and NIRISS imaging.  For the NIRCam field, this is similar to the comparison presented in \citet[][see their Fig.~9]{ers_overview}, although our photometry has since been improved on several fronts as described in \citet{ers_dolphot}.  The rich population of young stars ($<$1 Gyr) in the NIRCam field also reveals good agreement in the locations of the blue core helium burning (CHeB) sequence, seen extending from \textbf{($m_{F090W}-m_{F150W}$,$m_{F090W}$)}$\approx$(0.6,23.0) to (0.1,21.3) and the red CHeB sequence from \textbf{($m_{F090W}-m_{F150W}$,$m_{F090W}$)}$\approx$(0.9,22.1) to (1.2,19.8).  For the NIRISS field, we provide the first comparison between isochrones and photometry in the F090W and F150W filters.  Again, there is good agreement over the various phases of stellar evolution present, including the main sequence \textbf{($m_{F090W}-m_{F150W}$,$m_{F090W}$}$\approx$ 0.5,28.5), the subgiant branch \textbf{($m_{F090W}-m_{F150W}$,$m_{F090W}$}$\approx$ 0.6,27.1) and the red giant branch (0.7$\gtrsim$\textbf{$m_{F090W}-m_{F150W}$}$\gtrsim$1.4, 20.7$\gtrsim$\textbf{$m_{F090W}$}$\gtrsim$26.5), as seen in the NIRCam field.

We compare variations in photometric completeness across the NIRCam and NIRISS fields in the bottom row of Fig.~\ref{jwstcmd_fig}.  Minor differences between NIRCam and NIRISS in signal-to-noise ratio and completeness, even for identical exposure times, are expected based on non-identical throughput, spatial resolution and available readout modes.  Hence, the NIRISS observations saturate at fainter magnitudes than the NIRCam observations.  For our WLM imaging, we also see that completeness in the more densely populated NIRCam field is driven primarily by crowding.  By dividing each of the NIRCam and NIRISS fields into equally-populated bins of R$_{SMA}$, we see a clear spatial trend within the NIRCam field such that the less crowded outer bins are complete to fainter apparent magnitudes.  On the other hand, the more distant NIRISS field is sufficiently sparse that its completeness limits are insensitive to distance from the center of WLM, and this reduced crowding yields 50\% completeness limits that are 0.4-1.2 mag deeper than any of the NIRCam fields.  In any case, the range of 50\% completeness limits seen across the radial bins for each instrument comfortably brackets the full-field values reported in \citet[][see their Table 4]{ers_dolphot}.

\subsubsection{Archival HST Photometry}

\begin{figure}
\gridline{\fig{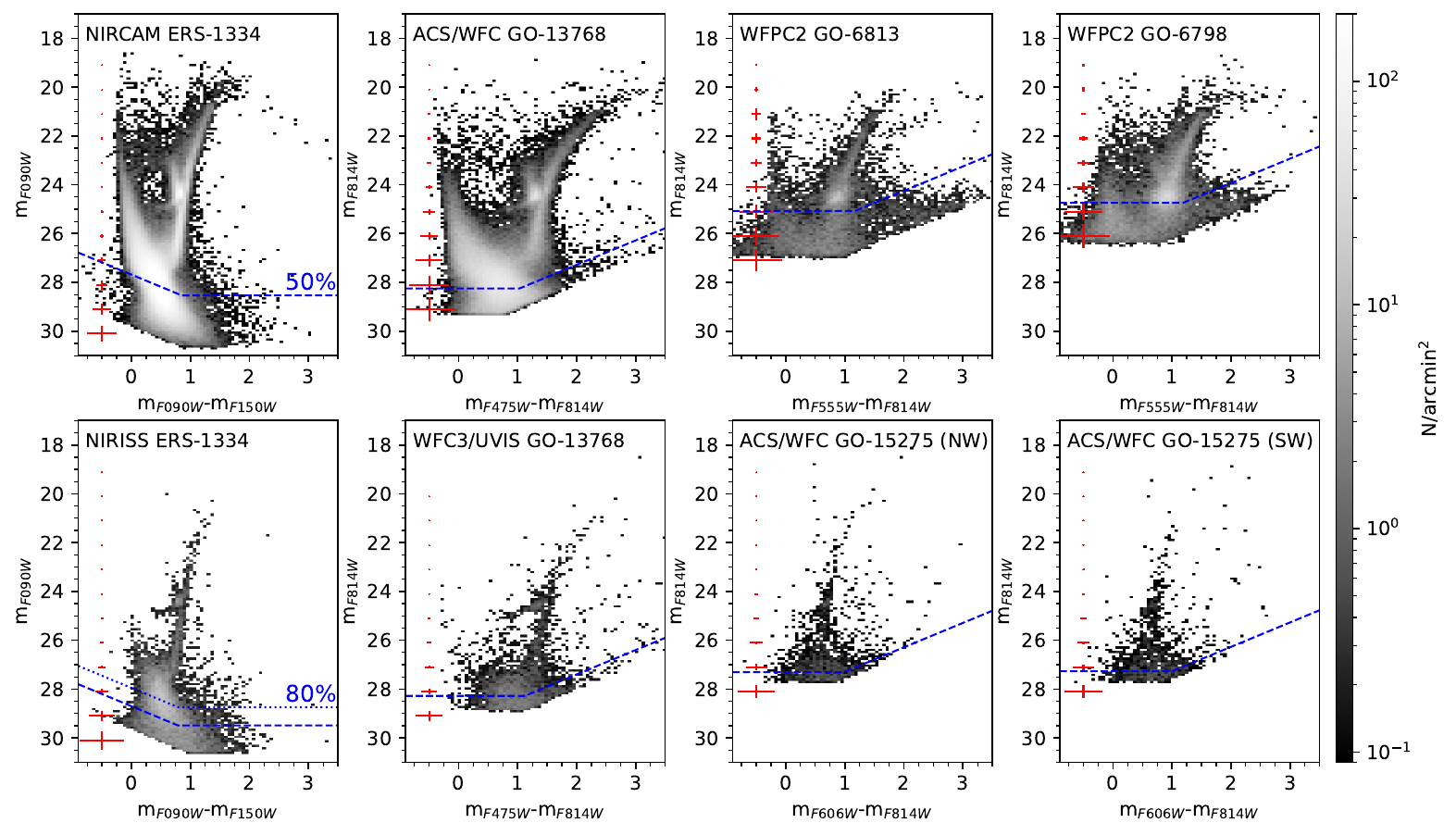}{0.99\textwidth}{}}
\caption{CMDs for each of the individual fields included in our spatially resolved SFH analysis, greyscaled by projected stellar density as indicated by the color bar.  Median photometric errors in magnitude bins are indicated in red, and the 50\% faint completeness limits are indicated by a dashed blue line in each panel.  For the NIRISS field (lower \textbf{left} panel), we also indicate the 80\% completeness limit used for SFH fitting using a dotted blue line (see Sect.~\ref{sfhfitsect}).  Axis scaling is kept the same across all panels to highlight differences in photometric depth across fields.} 
\label{allcmd_fig}
\end{figure}

CMDs of all of the HST fields we analyze are shown in Fig.~\ref{allcmd_fig}, where the NIRCam and NIRISS fields are also included for comparison.  To illustrate the range of photometric depths across all of the fields, we have used the same axis scaling in all cases, with the $I$-band equivalent filter of each instrument on the vertical CMD axis.  In the top row of Fig.~\ref{allcmd_fig}, we show the innermost four fields (R$_{SMA}$$\lesssim$R$_{hl}$), while the four outer fields (R$_{SMA}$$\gtrsim$2R$_{hl}$) are shown in the bottom row.  The shallowest fields are the WFPC2 fields (top row, right), severely impacted by crowding and ultimately providing only coarse constraints on the spatially resolved SFH (see Sect.~\ref{radgradsect}).  The outermost ACS fields (bottom row, right) benefit from both reduced crowding and improved throughput, extending $>$1 mag fainter than the WFPC2 fields despite shorter exposure times (see Table \ref{obstab}).  Meanwhile, thanks to the unique imaging capabilities provided by JWST, our NIRCam and NIRISS imaging extends even deeper than the deepest HST imaging despite shorter on-sky exposure times (also see \citealt{ers_dolphot}).  

Photometry of several of the HST fields in our sample has been presented previously \citep{dolphin_wlm,albers19}, so we have reprocessed these fields with \texttt{DOLPHOT} to ensure a homogeneous reduction and analysis strategy.
We also note that one of the WFPC2 fields (GO-6813, PI: Hodge) targets the only known globular cluster in WLM.  Since we are concerned with trends in the field population of WLM, we excised from our catalogs all sources (and artificial stars) within the cluster's tidal radius (31$\arcsec$; \citealt{hodge99}), while a recent reanalysis of the cluster can be found in \citet{sarajedini23}.  

\section{Analysis \label{anasect}}

Our primary goal is to measure trends in the lifetime SFH of WLM as a function of (semi-major axis equivalent) distance from its center R$_{SMA}$.  In addition, the availability of fields at multiple position angles over a range of R$_{SMA}$ provides the opportunity to test for any azimuthal dependence of the resultant radial age trends.  To accomplish these goals, we divided the field of view of each pointing into equally populated \enquote{radial bins} (in practice, elliptical annuli).  The radial bins we used are shown using colored lines in Fig.~\ref{obs_fig}, and were set by first excluding overlapping spatial regions from the shallower of any two overlapping fields, and then sorting the stars in each pointing by their observed R$_{SMA}$.  Each radial bin of each field then functions as a statistically independent stellar sample to which we fit a SFH (detailed below).  The number of radial bins per field was a necessary compromise between spatial resolution, needed to measure radial stellar age gradients within WLM, versus number statistics, which drive statistical uncertainties on SFH fits.   We used four radial bins for the deep, well-populated NIRCam and ACS fields, and two radial bins for the remainder of the fields, although the SFH trends we measure (see Sect.~\ref{resultsect}) were ultimately insensitive to reasonable (factor of 2) variations in the number of chosen radial bins. 

\subsection{Star Formation History Fitting \label{sfhfitsect}}

We fit SFHs to CMDs of resolved stars in each radial bin using the software package \texttt{MATCH} \citep{match}.  To calculate the best fitting SFH, \texttt{MATCH} compares the observed CMD to synthetic photometry produced by linear combinations of simple stellar populations (SSPs) at user-selected age intervals, seeking the best fit using a Poisson maximum likelihood statistic.  Our input assumptions for SFH fitting are based on the vast database of lifetime SFH results for Local Volume dwarfs using \texttt{MATCH} \citep[e.g.,][]{mcquinn10,weisz11,weisz14,mcquinn18,savino23}, and identical in practice to those of \citet{ers_kristy}, so here we summarize them briefly.  

For each radial bin, we perform two separate SFH fits using either the PARSEC \citep{parsec} or Bag of Stellar Tracks and Isochrones (BaSTI; \citealt{basti}) libraries of stellar evolutionary models, assuming solar-scaled abundance ratios.  These are currently the only libraries of stellar evolutionary models that have been updated to include the latest (Sirius-based) photometric calibration in the native filter system of NIRCam.  For NIRISS, updated bolometric corrections are not yet available in \texttt{MATCH}, so we used photometric zeropoints and filter transmission curves from the equivalent (F090W and F150W) NIRCam filters, which yield similar colors to $\lesssim$0.01 mag at fixed \textbf{$m_{F090W}$} for stars with $T_{\rm eff}$$\geq$3000K.  Since the CMD bins used for SFH fitting have a width of 0.05 mag in color and 0.10 in magnitude (see below), such an offset has a negligible impact on our SFH fitting results.  Furthermore, if we were to exclude the NIRISS field from our sample entirely, the position-angle-dependent difference in radial age gradients we report would become more (rather than less) pronounced (see Sect.~\ref{radgradsect}).

For SFH fitting, we used time bins covering stellar ages of 6.6$\leq$Log$_{\rm 10}$(Age/yr)$\leq$10.15, with width $\delta$Log$_{\rm 10}$(Age/yr)$=$0.1 for lookback times of Log$_{\rm 10}$(Age/yr)$\leq$9 and $\delta$Log$_{\rm 10}$(Age/yr)$=$0.05 otherwise.  We assumed a \citet{kroupa} initial mass function, a binary fraction of 0.35 with a flat mass ratio distribution and a metallicity spread of $\Delta$[M/H]$=$0.15 dex in each time bin.  We required metallicity to increase monotonically with time from an initial metallicity $-$2.0$\leq$[M/H]$\leq$$-$1.4 to a present-day metallicity $-$1.0$\leq$[M/H]$\leq$$-$0.5.  These initial and final metallicity ranges are based on independent measurements from RR Lyrae, red giants, early-type supergiants, and HII regions (see Sect.~\ref{wlmtestsect}) 
and function only as loose constraints (i.e., taking systematic offsets in abundance scales into consideration).  Accordingly, the choice to restrict the allowed metallicity ranges during SFH fitting did not affect our results beyond their uncertainties. In fact, for the most well-populated and/or deepest (NIRCam, NIRISS and ACS) fields, the best-fit age-metallicity relations fell within the assigned range without enforcing any \textit{a priori} constraints at all.  

Our fields cover a substantial baseline of distance from the center of WLM, so in principle, enforcing identical initial and final metallicity ranges over fields at different R$_{SMA}$ could bias our results in the case of a radial metallicity gradient.  However, spectroscopy of red giants in WLM out to $>$2R$_{hl}$ along both the major and minor axes indicates that any such gradient is very mild and statistically consistent with null at 1-$\sigma$ \citep{leaman13,taibi22}.  Our range of allowed present-day metallicities is also consistent with abundances reported for nine HII regions in WLM \citep{skillman89,lee05}, noting that these abundances show no trend with R$_{SMA}$, although the sampled HII regions extend to only R$_{SMA}$$\lesssim$2.7$\arcmin$ ($<$0.6R$_{hl}$).  In summary, we apply identical restrictions to the allowed ranges of initial and final metallicities for all radial bins of all fields regardless of their R$_{SMA}$.  We also reiterate that while we have placed empirically motivated constraints on the initial and final [M/H] values that are applied identically to all radial bins, the exact form of the age-metallicity relation is allowed to vary on a per-radial-bin basis.

For SFH fitting, we assume a distance modulus of $(m-M)_{0}$$=$24.93 mag (D$_{\odot}$$=$968 kpc) based on a fit to the tip of the red giant branch as in \citet{albers19} and \citet{ers_kristy}, consistent with other recent measurements \citep{mcconnachie05,jacobs09,anand21,lee21}.  We assume 
a foreground extinction of A$_{V}$=0.10 mag based on the \citet{sf11} recalibration of the \citet{sfd} extinction maps, and a \citet{fitzpatrick99} extinction law with $R_{V}$$=$3.1.   While \texttt{MATCH} is capable of including additional internal differential extinction in SFH solutions, we chose not to include such additional extinction in our fits.  We found that zero differential extinction was the statistically preferred solution for the deep ACS field, imaged in the most extinction-sensitive filter combination across our sample (\textbf{$m_{F475W}-m_{F814W}$}).  This result is consistent with the lack of differential extinction reported by \citet{albers19}, in agreement with evidence for the lack of internal extinction in gas-rich dwarfs at low (i.e., WLM-like) metallicities more generally \citep{mcquinn15}.\footnote{An obvious exception is the SMC, showing substantial spatially variable differential extinction \citep[e.g.,][]{cignoni13,skowron21}.  However, the SMC has a present-day (H$\alpha$-based) star formation rate over an order of magnitude higher than WLM \citep[e.g.,][]{k18} and is dramatically impacted by prolonged interactions with the LMC (see Sect.~\ref{obscompsect}).}  Our fits also include a Galactic foreground component with a color-magnitude distribution determined from the TRILEGAL galaxy model \citep{trilegal}, although foreground contamination at the galactic latitude of WLM ($B=-73.63^{\circ}$) is predicted to be minimal ($\lesssim$3 sources/arcmin$^{2}$ over the sampled CMD regions), so our results are insensitive to the inclusion of this component.  

When comparing observed and synthetic CMDs, \texttt{MATCH} divides each CMD into bins of width 0.05 mag in color and 0.10 mag in magnitude.  For each radial bin, this comparison is performed over the entire CMD down to the 50\% completeness limit in each of the two filters, ascertained using cospatial artificial stars (i.e., located in the same radial bin).  This strategy has the advantage of exploiting the improved photometric depth in radial bins with lower projected source densities (both observed and intrinsic), particularly relevant to the more crowded inner fields (e.g., the bottom row of Fig.~\ref{jwstcmd_fig}).  Statistical uncertainties on the best-fit SFH are calculated using a Hybrid Monte Carlo approach detailed in \citet{dolphin_randerr}, which uses Hamiltonian dynamics to efficiently sample high-dimensional parameter spaces \citep{duane87}.

For the NIRISS field, our photometry 
is $\gtrsim$1 mag deeper than any of our other fields (e.g., Figs.~\ref{jwstcmd_fig} and \ref{allcmd_fig}), and also has a relatively low projected density of WLM members due to its more external location (compared, for example, to the NIRCam field; see Figs.~\ref{obs_fig}-\ref{jwstcmd_fig}).  As a result, the $(m_{F090W}-m_{F150W})$ color of background galaxies (e.g., Fig.~1 of \citealt{warfield23}) conspires with their increased frequency at fainter apparent magnitudes (e.g., Table 5 of \citealt{rafelski15}) to contaminate the main sequence of the NIRISS field.  Therefore, we chose to move our faint limit for SFH fitting in the NIRISS field brightward to the 80\% (rather than 50\%) faint completeness limits to mitigate background galaxy contamination.  However, we found that either excluding the CMD region occupied by background galaxies from SFH fitting and/or using alternate choices of faint completeness limits ultimately did not affect our results in Sect.~\ref{resultsect} beyond their uncertainties. 

\begin{figure}
\gridline{\fig{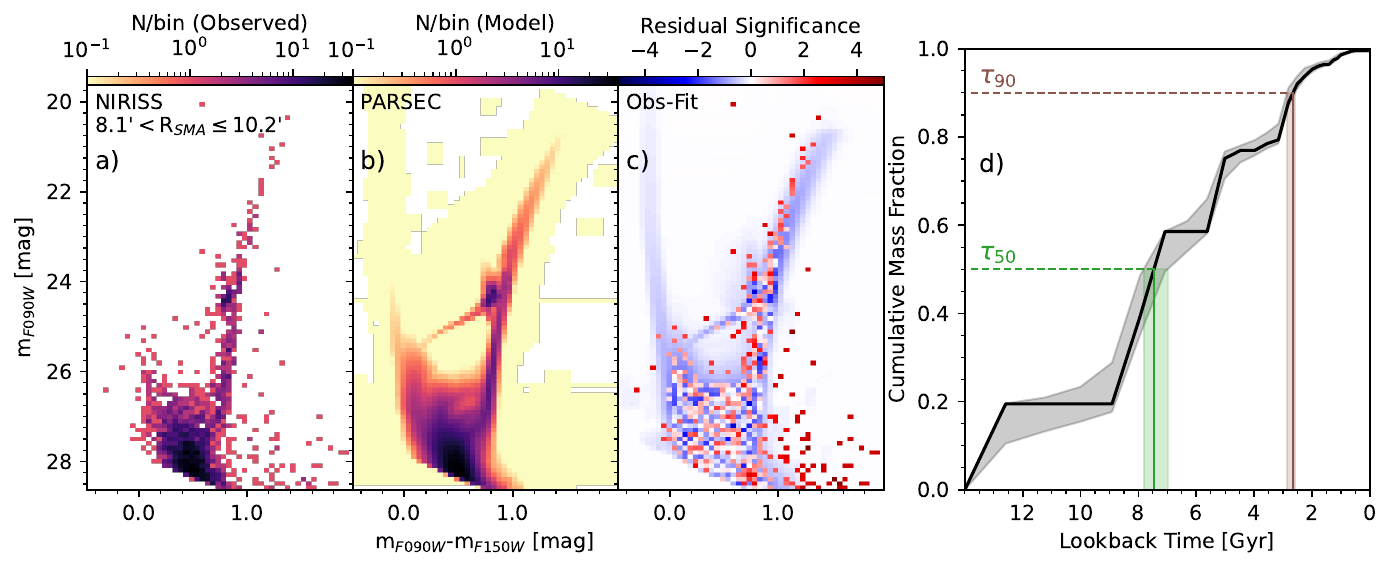}{0.99\textwidth}{}}
\caption{An example SFH fit to the inner radial annulus of the NIRISS field.  The three Hess diagrams on the left illustrate, from left to right: the observed (a) and modeled (b) star counts per color-magnitude bin, and the fit residuals (c) in terms of Poisson significance in the sense (data-model).  The right-hand plot (d) shows the best-fit cumulative distribution of stellar mass formed as a function of lookback time (black) and its statistical uncertainty (grey shading).  The metrics we use to quantify radial stellar age gradients, $\tau_{50}$ and $\tau_{90}$, correspond to the lookback times by which 50\% and 90\% of the cumulative stellar mass had been formed.  The values and uncertainties of $\tau_{50}$ and $\tau_{90}$ are obtained via interpolation in the cumulative stellar mass distribution and its $\pm$1-$\sigma$ uncertainty envelope.} 
\label{example_sfh_fig}
\end{figure}

An example of our SFH fitting procedure is shown in Fig.~\ref{example_sfh_fig} for the inner radial bin of the NIRISS field, assuming PARSEC evolutionary models for SFH fitting.  A map of the fit residuals over the CMD (Fig.~\ref{example_sfh_fig}c) functions as a useful diagnostic of SFH fit quality, illustrating the difference between the observed and best-fit number of stars in each CMD bin in terms of Poisson significance.  For a near-perfect SFH fit, these residuals would be uncorrelated with CMD location.  For the NIRISS field, the fit quality is generally very good, modulo minor offsets in three locations.  First, discrepancies are seen near the red clump and horizontal branch, advanced stages of stellar evolution that still suffer from modeling uncertainties \citep[e.g.,][]{gallart05}, although we found that excluding these CMD regions from SFH fitting had an essentially negligible ($<$7\%) impact on the best-fit CSFHs.  Second, color offsets of $\lesssim$0.05 mag on the upper red giant branch are similar to values previously reported for near-infrared broadband filters, including those used here \citep{cohen15,ers_overview,ers_kristy,mcquinn24b,garling24}, and may be related to the use of a fixed mixing length parameter in stellar models (see \citealt{joyce23} for a review).  Third, the reddest sources ($m_{F090W}-m_{F150W}$$\gtrsim$1.2) are unresolved background galaxies that appear sufficiently point-source-like to pass our photometric quality cuts.

To quantify spatial variations in the SFH of WLM, we use metrics based on the cumulative star formation history (CSFH).  Using the CSFH output by \texttt{MATCH}, we measure the lookback times by which 90\% and 50\% of the cumulative stellar mass had been formed, denoted $\tau_{90}$ and $\tau_{50}$.  These CSFH-based metrics are illustrated for our example NIRISS radial bin in Fig.~\ref{example_sfh_fig}d, and have two advantages: First, the use of a CSFH rather than a binned histogram of star formation rate in each time interval mitigates the impact of correlated star formation rate uncertainties that can be sensitive to the chosen time binning scheme, potentially yielding large star formation rate uncertainties in neighboring time bins.  Conversely, uncertainties in $\tau_{90}$ and $\tau_{50}$ are calculated straightforwardly by interpolating in the CSFH and its uncertainty envelope.  Second, by using a cumulative distribution, $\tau_{50}$ and $\tau_{90}$ values are sensitive only to the shape of the star formation history, and insensitive to its normalization (which can depend, for example, on the assumed initial mass function).  

\subsection{Dependence on Assumed Stellar Evolutionary Models \label{modelcompsect}}

\begin{figure}
\gridline{\fig{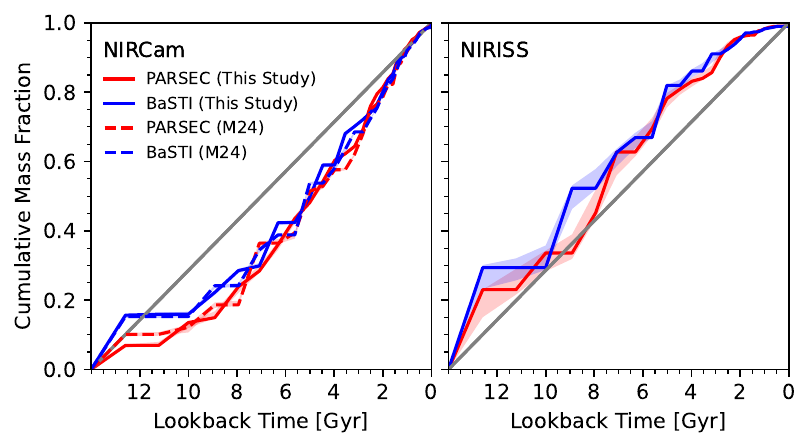}{0.9\textwidth}{}}
\caption{Comparison between best-fit CSFHs assuming PARSEC (red) or BaSTI (blue) stellar evolutionary models, with statistical uncertainties indicated by shading.  The left panel shows results for the NIRCam field, obtained by statistically combining SFH fits in individual radial bins.  For comparison, CSFHs from \citet{ers_kristy} are overplotted as \textbf{dashed} lines.  The right panel shows CSFHs for the NIRISS field, and the grey line in each panel corresponds to a constant SFH.} 
\label{csfh_models_fig}
\end{figure}

In Fig.~\ref{csfh_models_fig}, we compare CSFHs obtained assuming PARSEC and BaSTI evolutionary models for NIRCam and, for the first time, NIRISS imaging.  These full-field CSFHs for each instrument were calculated by statistically combining SFH fitting results for each radial bin (for further details see \citealt{weisz11,dolphin_syserr}).  
Because different evolutionary libraries tend to make similar assumptions in their methodology (e.g., use of mixing length theory to treat convection), model-to-model differences in SFH results should be viewed as lower limits on the impact of changes in input physics.\footnote{A prescription for calculating the systematic uncertainty contribution to SFHs due to differences in assumed stellar evolutionary libraries was provided by \citet{dolphin_syserr}.  However, this prescription requires the direct comparison of SFH results from more than two libraries of stellar evolutionary models to calculate standard deviations, while updated NIRCam photometric calibrations are only currently available for the PARSEC and BaSTI libraries.}    
Nevertheless, such a comparison provides a useful empirical estimate.  For our fits to the NIRCam and NIRISS fields, Fig.~\ref{csfh_models_fig} demonstrates good agreement between CSFHs fit assuming PARSEC and BaSTI evolutionary models, particularly at more recent lookback times where the impacts of model-to-model variations tend to be relatively smaller (e.g., Appendix B of \citealt{albers19}).  While the BaSTI models estimate a higher stellar mass fraction formed at ancient times compared to the PARSEC models for the NIRCam and NIRISS fields, our data are insufficient to assess whether this feature is due to differences in the model input physics and is therefore universal.  However, for the lookback times we examine here (i.e., corresponding to $\tau_{50}$ and $\tau_{90}$), evidence within our sample suggests that this is unlikely to be the case:  Comparing BaSTI- and PARSEC-based $\tau_{50}$ values across all radial bins in all fields, the difference of the two $\Delta$$\tau_{50}$(BaSTI$-$PARSEC) has a (weighted) mean and standard deviation of 0.19 and 0.24 Gyr respectively.

The left panel of Fig.~\ref{csfh_models_fig} illustrates that our NIRCam CSFHs are also in good agreement with the SFHs presented by \citet{ers_kristy}, fit to the full field of view using the same two stellar evolutionary models and an earlier generation of \texttt{DOLPHOT} photometry (presented in \citealt{ers_overview}), and (although not shown here) results obtained using the same data and PARSEC evolutionary models but a different SFH fitting code \citep{garling24}.  We therefore present results assuming PARSEC models, noting that radial age gradient slopes measured assuming BaSTI models are consistent to within 1-$\sigma$ statistical uncertainties and are also provided for comparison (see Sect.~\ref{radgradsect}).   

It is clear from Fig.~\ref{csfh_models_fig} that, regardless of assumed stellar evolutionary model, the more externally located NIRISS field is substantially older on average than the NIRCam field (consistent with the CMDs in Fig.~\ref{jwstcmd_fig}).  A qualitatively similar difference in SFHs between inner and outer fields in WLM was found comparing the UVIS and ACS fields \citep{albers19}, so we turn to our CSFHs to measure radial age trends in WLM quantitatively. 

\section{Results and Discussion\label{resultsect}}
\subsection{Spatially Resolved Stellar Age Trends in WLM \label{radgradsect}}

\begin{figure}
\gridline{\fig{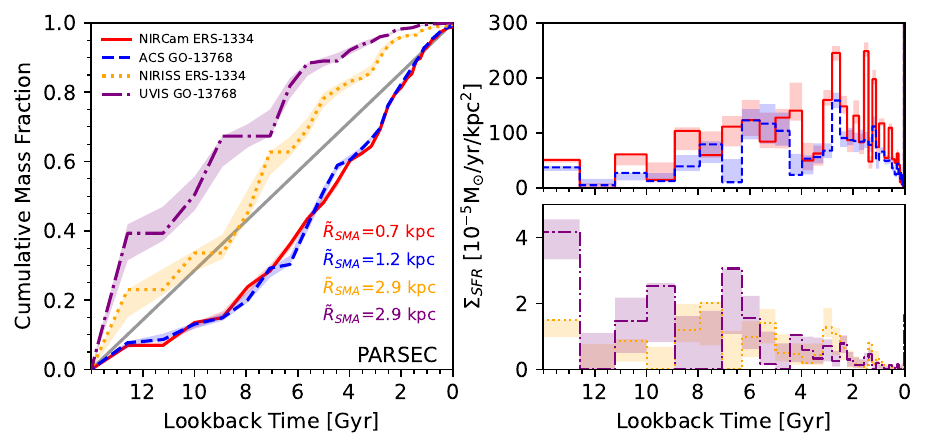}{0.99\textwidth}{}}
\caption{\textbf{Left:} Comparison between the CSFHs of our deep JWST and HST fields.  The median R$_{SMA}$ of each field is given in the lower right corner.  The more externally located UVIS and NIRISS fields, at the same R$_{SMA}$ but different orientations, both assemble their mass earlier than the ACS and NIRCam fields, but have non-identical CSFHs. Shading indicates statistical uncertainties calculated following \citet{dolphin_randerr}. \textbf{Right:} Star formation rate density $\Sigma_{\rm SFR}$ versus time for each of the four fields shown in the left panel.  Two separate y-axis scales are used due to the lower values of $\Sigma_{\rm SFR}$ in the more external UVIS and NIRISS fields by $\sim$2 orders of magnitude.
\label{csfhdeep_fig}}
\end{figure}

The four fields in our sample with the deepest imaging are the NIRCam, NIRISS and GO-13768 ACS and UVIS pointings, with photometry extending to the ancient main sequence turnoff (or deeper; see Figs.~\ref{jwstcmd_fig}$-$\ref{allcmd_fig}).
Together, these fields sample a range of both R$_{SMA}$ as well as position angle relative to the center of WLM (see Fig.~\ref{obs_fig}).  In the left panel of Fig.~\ref{csfhdeep_fig}, we directly compare the CSFHs of these four fields, exploiting the small statistical uncertainties on their CSFHs provided by deep imaging.   
The NIRCam and ACS fields (R$_{SMA}$$\lesssim$R$_{hl}$), which are spatially overlapping, have nearly identical CSFHs to $\leq$5\%, consistent with \citet{ers_kristy}.  
However, a comparison between the more external UVIS and NIRISS fields, which lie at identical R$_{SMA}$ but different position angles, 
suggests that the spatially resolved SFH of WLM may vary azimuthally at fixed R$_{SMA}$.  

A correlation between lifetime SFH and position angle in WLM is particularly tantalizing in light of the recent suggestion that WLM is undergoing ram pressure stripping. 
In particular, \citet{yang22} report the detection of H\textsc{I} clouds asymmetrically distributed on the trailing side of WLM, opposite its direction of motion, which they posit are a consequence of ram pressure stripping by the intergalactic medium (IGM; we return to this point in more detail below in Sect.~\ref{causesect}). 
To quantify whether radial trends in the SFH of WLM show any dependence on orientation relative to its direction of motion, we divided our fields into two samples, a \enquote{leading edge} and \enquote{trailing edge} sample, selected based on position angle relative to the direction of WLM's proper motion (indicated with an arrow in Fig.~\ref{obs_fig}).  The elliptical annuli included in the leading edge and trailing edge samples are shown 
in Fig.~\ref{obs_fig} using magenta-blue and black-yellow color coding respectively.  The annuli selected for inclusion in either the leading or trailing edge samples were chosen to cover a similar baseline of R$_{SMA}$ but essentially opposite orientations relative to the proper motion of WLM insofar as the placement of available imaging allows.   
Because the center of WLM lies within the NIRCam field (by design), this was the only field that was spatially subdivided into both leading and trailing edge subsamples based on position angle, each with their own set of radial bins that were analyzed independently (shown in Fig.~\ref{obs_fig}; note that these differ from the radial bins shown in the inset of Fig.~\ref{jwstcmd_fig}, which lack any position angle requirements).  In addition, due to their position angles, the two GO-15275 ACS fields and the GO-6798 WFPC2 field were not included in either the leading or trailing edge samples, and the locations of their elliptical annuli (which we retain in our analysis to calculate galaxy-wide radial age gradients below) are shown in light and dark green in Fig.~\ref{obs_fig}.

\begin{figure}
\gridline{\fig{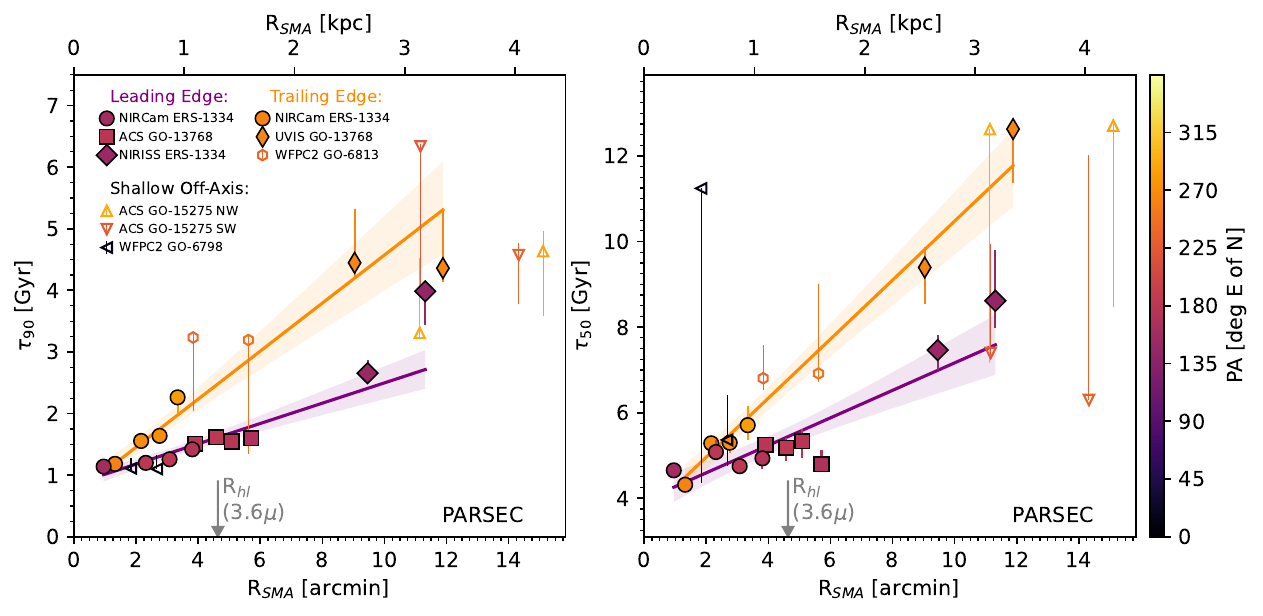}{0.99\textwidth}{}}
\caption{The lookback times to form 90\% (left) and 50\% (right) of the cumulative stellar mass for each individual radial bin of each field (per-field radial bin locations are indicated by the curved lines in Fig.~\ref{obs_fig}) as a function of R$_{\rm SMA}$.  Fields are color-coded by median position angle east of north.  Fields along the leading edge of WLM are preferentially younger than fields on the trailing edge, and have a shallower radial age gradient measured using both $\tau_{90}$ and $\tau_{50}$.  Maximum likelihood linear fits are shown using colored lines with 1-$\sigma$ (16th-84th percentile) uncertainties indicated by shading.  Results shown here assume PARSEC stellar evolutionary models, and fit coefficients including alternate stellar evolutionary models are given in Table \ref{coefftab}, producing essentially identical results.  The arrow indicates the half-light radius of WLM R$_{hl}$ used to compare our observed radial stellar age gradients with simulation predictions (see Sect.~\ref{compsimsect}).} 
\label{radgrad_fig}
\end{figure}

Within each sample (leading or trailing edge), we leverage the available baseline of R$_{SMA}$ to measure radial stellar age gradients.  The age gradient is quantified by fitting a line to the per-radial-bin values of $\tau_{90}$ and $\tau_{50}$ versus R$_{SMA}$ to obtain the slopes $\delta$$\tau_{90}$/$\delta$R$_{SMA}$ and $\delta$$\tau_{50}$/$\delta$R$_{SMA}$.  The best-fit coefficients and their uncertainties were obtained from posterior distributions estimated using the affine-invariant Markov Chain Monte Carlo sampler implemented in the  
\texttt{emcee} package \citep{emcee}, using 50 walkers with an initial burnin of 500 interations followed by 5000 production iterations (sufficient given autocorrelation lengths of $<$50 iterations).  The coefficients of the linear fits are provided in Table \ref{coefftab}, and the best-fit lines (and their 1-$\sigma$ uncertainties) are shown for the leading and trailing edge samples in Fig.~\ref{radgrad_fig}.
We find that 
the fields towards the leading edge of WLM are preferentially younger than those towards the trailing edge, with a shallower radial age gradient, and potential explanations are discussed below in Sect.~\ref{causesect}.  

For comparison to both simulations and observations of other dwarfs, we also provide in Table \ref{coefftab} a single set of galaxy-wide values for radial stellar age gradients in WLM.  These fits exploit the full extent of available spatial coverage, including the three shallow fields excluded from the leading and trailing edge samples (shown as open triangles in Fig.~\ref{radgrad_fig}), and include radial bins selected from the full NIRCam field of view without position angle restrictions imposed (as shown in the inset in the lower right-hand panel of Fig.~\ref{jwstcmd_fig}). 

\begin{deluxetable}{lcccc}
\tablecaption{Coefficients of Linear Fits to Radial Stellar Age Gradients \label{coefftab}}
\tablehead{
\colhead{Sample} & \colhead{$\delta$$\tau_{90}$/$\delta$R$_{SMA}$} & \colhead{$\tau_{90}$(R$_{SMA}$=0)} & $\delta$$\tau_{50}$/$\delta$R$_{SMA}$ & \colhead{$\tau_{50}$(R$_{SMA}$=0)} \\ \colhead{} & \colhead{Gyr/kpc} & \colhead{Gyr} & \colhead{Gyr/kpc} & \colhead{Gyr}}
\startdata
\multicolumn{5}{c}{Assuming PARSEC Evolutionary Models} \\
\hline
Leading Edge & 0.58$^{+0.13}_{-0.13}$ & 0.86$^{+0.14}_{-0.14}$ & 1.14$^{+0.32}_{-0.32}$ & 3.96$^{+0.41}_{-0.41}$\\
Trailing Edge & 1.40$^{+0.28}_{-0.26}$ & 0.67$^{+0.22}_{-0.22}$ & 2.46$^{+0.32}_{-0.36}$ & 3.56$^{+0.39}_{-0.36}$\\
All Fields & 0.82$^{+0.10}_{-0.10}$ & 0.76$^{+0.10}_{-0.11}$ & 1.60$^{+0.23}_{-0.22}$ & 3.85$^{+0.29}_{-0.31}$\\
\hline
\multicolumn{5}{c}{Assuming BaSTI Evolutionary Models} \\
\hline
Leading Edge & 0.80$^{+0.14}_{-0.14}$ & 0.62$^{+0.14}_{-0.15}$ & 1.27$^{+0.22}_{-0.22}$ & 3.74$^{+0.32}_{-0.32}$\\
Trailing Edge & 1.40$^{+0.22}_{-0.20}$ & 0.62$^{+0.19}_{-0.20}$ & 1.97$^{+0.27}_{-0.32}$ & 3.96$^{+0.37}_{-0.34}$\\
All Fields & 0.78$^{+0.07}_{-0.07}$ & 0.74$^{+0.10}_{-0.09}$ & 1.23$^{+0.16}_{-0.16}$ & 4.16$^{+0.21}_{-0.22}$\\
\enddata
\end{deluxetable}

\subsection{Causes of Azimuthally-Dependent Radial Age Gradients \label{causesect}}

This is the first instance in which a position-angle-dependent difference in radial stellar age gradient slopes has been quantified in a dwarf galaxy.  Star formation that is asymmetrically distributed has been inferred from SFH fits to resolved CMDs in other cases, but only in galaxies showing highly irregular optical morphologies not seen in WLM.  These consist of blue compact dwarfs hosting a broad range of stellar ages \citep[e.g.,][]{annibali13,sacchi21} and the Magellanic Clouds, which recently collided with each other and have been interacting for Gyrs \citep[][and references therein]{cohen24,cohen24b}.  WLM, on the other hand, is an isolated dIrr that was likely even more isolated in the past, perhaps previously located near the Local Void \citep[e.g.,][]{tully19} based on its space velocity (\citealt{battaglia22,bennet24}; see the discussion by \citealt{yang22}).  Therefore, orientation-based trends in its age gradients, which sample lookback times as far back as $\sim$8 Gyr in the case of $\tau_{50}$ (see Fig.~\ref{radgrad_fig}), cannot be ascribed to past interactions.  Remaining explanations for the younger ages and flatter age gradient towards its leading edge consist of at least three non-exclusive possibilities.

The most mundane explanation for the azimuthally-dependent stellar age trends in WLM is that it is a result of line-of-sight projection effects, especially since the trailing edge sample lies along the minor axis while the leading edge sample does not.  The outermost fields in our sample (extending to $\gtrsim$3R$_{hl}$) are fortuitously positioned opposite the direction of WLM's motion and also off of the minor axis sampled by our trailing edge fields, but are too sparsely populated to firmly confirm or refute this hypothesis.

As a second possibility, filter- and instrument-dependent systematic offsets could, in principle, contribute to differences in the best-fit SFH at fixed R$_{SMA}$.  However, the exquisite agreement in both $\tau_{50}$ and $\tau_{90}$ between the NIRCam trailing edge field and the neighboring ACS field seen in Fig.~\ref{radgrad_fig} at R$_{SMA}$$\approx$3.9$\arcmin$ argues strongly against this possibility.  These results are fully consistent with \citet{ers_kristy}, who performed a direct comparison between NIRCam- and ACS-based SFHs for the region of spatial overlap, also finding excellent agreement.  Furthermore, the difference in radial age gradient slopes between the leading and trailing edge samples are seen clearly within the NIRCam field alone, so they cannot be entirely an artifact of observational systematics.

A third potential explanation for the more recent values of $\tau_{50}$ and $\tau_{90}$ towards the leading edge is that ram pressure 
has preferentially triggered star formation on the leading edge.  Despite the relative present (and likely past) isolation of WLM, there are both observational and theoretical lines of evidence that such ram pressure may be related to its proximity to the Milky Way and M31.  Based on the H\textsc{I} properties of Local Group dwarfs and their inferred past orbits, \citet{putman21} suggest that a Local Group medium is capable of stripping gas from dwarfs well beyond the virial radii of their massive hosts.  
Furthermore, stripping by a Local Group medium was suggested as a cause of the asymmetrical H\textsc{I} morphology of the Pegasus dIrr (\citealt{mcconnachie07}, but see \citealt{kniazev09}).  Simulations also indicate that environmental processes can impact star formation in dwarfs out to at least twice the virial radii of their hosts \citep{fillingham18}, with some models predicting correlations between $\tau_{90}$ and host distance out to several Mpc \citep{christensen24}.    

Another potential source of ram pressure is the IGM, as proposed for WLM by \citet{yang22}.  In addition to the Local Group medium advocated by \citet{mcconnachie07} in the case of the Pegasus dIrr, ram pressure stripping by the IGM has been suggested previously for the dwarfs Holmberg II and Pisces A \citep{bureau02,bernard12,beale20}, but never cross-matched with position-angle-dependent SFH trends from resolved stars.  However, such a preferential orientation of star formation is observed for massive jellyfish galaxies \citep{roberts22} showing gaseous tails as a consequence of ram pressure stripping, and the clouds of neutral hydrogen trailing WLM produce a similar 
H\textsc{I} morphology \citep[][see their Fig.~1]{yang22} despite its lower mass and more isolated environment.  Furthermore, 
simulations suggest that ram pressure may play an important role for isolated dwarfs via \enquote{cosmic web stripping} in which their gas can be depleted via encounters with filaments \citep{benitezllambay13}, supported observationally by more effective gas removal in low-mass halos located close to such filaments \citep{hoosain24}.  Simulations by \citet{wright19} advocate for the IGM to impact SFH trends in dwarfs, finding that ram pressure stripping by streams of gas in the IGM can be responsible for a period of enhanced star formation followed by a pause (also see \citealt{pasha23}).  For WLM, a pause in star formation is observed after reionization at a lookback time of $\sim$12.5 Gyr, and is seen throughout the observed range of R$_{SMA}$ (\citealt{albers19,ers_kristy,garling24}, see Fig.~\ref{csfhdeep_fig}).  Among the simulated dwarfs analyzed by \citet{wright19}, those with \enquote{gappy} SFHs including a pause tend to be gas-rich at $z$$=$0  
similar to WLM (see Table \ref{proptab}).  Such post-reionization pauses have now been detected in three other Local Group dwarfs in addition to WLM (Leo A, Aquarius, and Leo P), all less massive but similarly gas-rich, although it remains unclear whether such SFH pauses are related to interactions with the IGM more generally (see the discussion by \citealt{mcquinn24b}).  

It is also possible that ram pressure stripping of WLM could originate from its circumgalactic medium (CGM) instead of (or in addition to) the IGM.  The CGM of WLM was tentatively detected by \citet{zheng19}, and simulations of CGM stripping of dwarf satellites predict H\textsc{I} column densities of order $\sim$10$^{19}$ atoms/cm$^{2}$ in their tails \citep{zhu24}, perhaps not incompatible with observations of the trailing HI clouds \citep{yang22}.  Furthermore, the wind tunnel simulations by \citet{yang22}, advocating for the IGM stripping scenario, required a relatively low stellar mass for WLM (M$_{\star}$$=$10$^{7}$M$_{\odot}$), a high H$\textsc{I}$ mass (M$_{\rm HI}$$=$10$^{8}$M$_{\odot}$), a high IGM density ($\geq$5.2$\times$10$^{-5}$ atoms/cm$^{3}$), and a high space velocity (300$-$500 km/s).  While the latter value is consistent with observations (\citealt{bennet24} estimate a space velocity of $\sim$400 km/s), the required IGM density is over an order of magnitude higher than that required to strip Holmberg II or the Pegasus dIrr (albeit with large uncertainties; \citealt{bureau02,mcconnachie07}), as high as the density needed to strip Milky Way dwarf satellites \textit{at perigalacticon} \citep{putman21}.  On the other hand, the \citet{zhu24} simulations predict that the conditions for CGM stripping are fairly specific both temporally and spatially, with CGM stripping occurring relatively rapidly ($\sim$500 Myr) when the galaxy is located near perigalacticon.  Future improvements in either the detection threshold of low-column-density gas and/or uncertainties on WLM's orbital history may ultimately be needed to distinguish between CGM and IGM stripping scenarios observationally.  

\subsection{Comparison to Observed Stellar Age Gradients in Local Volume Dwarfs \label{obscompsect}}

A comparison of our results against radial age gradients in other dwarfs is hampered by the lack of quantitative radial age gradient measurements currently available, motivating our analysis strategy for WLM here. 
Among the few available measurements, \citet{hidalgo13} examined radial age gradients in four isolated dwarfs.  For each of their targets, they measured 
$\tau_{10}$ and $\tau_{95}$ in radial bins, normalized to the exponential scalelength $h_{r}$.   
To perform a direct comparison with the \citet{hidalgo13} results, we calculated $\tau_{95}$ in each of our radial bins, finding gradients of $\delta$$\tau_{95}$/$\delta$R$_{SMA} = 0.53^{+0.15}_{-0.14}$ (0.67$^{+0.17}_{-0.15}$) Gyr/kpc for the WLM leading (trailing) edge fields. 
Assuming the exponential scalelength $h_{r}$$=$131.7$\arcsec$ for WLM from our ellipse fits, 
even the steeper trailing edge gradient corresponds to only $\sim$0.5 Gyr/$h_{r}$.  Comparing to the \citet{hidalgo13} sample, the Cetus dSph is the only one of the four dwarfs they analyzed that has such a flat radial age gradient.  Furthermore, while the four dwarfs targeted by \citet{hidalgo13} were selected based on their relative isolation, they sample a narrow range of both morphological type and stellar mass.  Their four targets are all dSphs or transition dwarfs, and are all more than an order of magnitude less massive than WLM \citep[e.g.,][]{mcconnachie12}.  Therefore, larger observational samples are needed to determine whether the radial gradients in $\tau_{95}$ observed in WLM and Cetus are unusual and test whether such gradients correlate with mass or morphology in isolated dwarfs\footnote{In the context of our results, it bears mention that the spatial coverage of Cetus (as well as Phoenix and, to a lesser extent, LGS3) available for the \citet{hidalgo13} study is azimuthally incomplete.  If these galaxies have position-angle-dependent age gradients as we report for WLM here, incomplete azimuthal sampling could bias measurements of their radial age gradients.}.  We refrain from a comparison with trends in $\tau_{10}$, hampered by larger uncertainty  
contributions at ancient lookback times. 

Radial stellar age gradients in the SMC were measured from CSFHs by \citet{cohen24b}, who found gradients that are outside-in, but with flatter slopes versus R$_{SMA}$, particularly for $\tau_{50}$.  Specifically, \citet{cohen24b} report $\delta$$\tau_{50}$/$\delta$R$_{SMA}$=0.82$^{+0.12}_{-0.16}$ (0.96$^{+0.13}_{-0.16}$) Gyr/kpc and $\delta$$\tau_{90}$/$\delta$R$_{SMA}$=0.61$^{+0.08}_{-0.07}$ (0.62$^{+0.09}_{-0.08}$) Gyr/kpc assuming PARSEC (BaSTI) models.  However, the SMC is over an order of magnitude more massive than WLM \citep[e.g.,][]{deleo24}, and is also an extreme case of an interacting galaxy.  Unlike WLM, the SMC exhibits a complex multi-component geometry with a significant line-of-sight depth likely due to numerous interactions with the more massive LMC over the last several Gyr \citep[e.g.,][and references therein]{besla12,sub17,murray24}.  In particular, the wing of the SMC deviates from the overall outside-in age gradient, harboring anomalously young stellar populations likely due to LMC-SMC interactions.  The SMC wing was therefore excluded from the \citet{cohen24b} fits, and both the young age of the wing (compared to other fields at similarly large R$_{SMA}$) as well as the overall outside-in age gradient in the SMC are consistent with previous studies of its spatially resolved SFH \citep[e.g.,][]{noel09,cignoni13}.

Our results may also be compared with radial stellar age trends in two Local Volume dwarfs analyzed by \citet{sacchi18,sacchi21}, NGC 4449 and UGC 4483.  For each of these targets, SFHs were fit to resolved CMDs of different spatial regions selected using isodensity contours, providing the opportunity to estimate radial age gradients.  The blue compact dwarf UGC 4483 has a present-day outside-in age gradient, with the SFHs fit to four radial regions independently by \citet{sacchi21} translating to age gradients of $\delta$$\tau_{50}$/$\delta$R$_{SMA}$$\approx$3 Gyr/kpc and $\delta$$\tau_{90}$/$\delta$R$_{SMA}$$\approx$1.5 Gyr/kpc, marginally steeper than WLM.  For NGC 4449, \citet{sacchi18} find a gradient in $\delta$$\tau_{50}$/$\delta$R$_{SMA}$ that is flat to within uncertainties, and a very shallow outside-in gradient of $\delta$$\tau_{90}$/$\delta$R$_{SMA}$$\approx$0.2 Gyr/kpc.  NGC 4449, which has a complex, irregular stellar and gaseous morphology, is currently undergoing tidal interactions with neighboring galaxies, and it is unclear whether such interactions are related to flat age gradients.  Interestingly, NGC 6822, the other nearby dwarf with flat radial age gradients (in $\tau_{75}$ and $\tau_{95}$; \citealt{cannon12,fusco14}), has multiple lines of evidence for past interactions based on its stellar morphology \citep{zhang21,tantalo22}.  While simulations predict that radial age gradient slopes range from outside-in to flat fairly continuously (see Sect.~\ref{compsimsect} below), spatially resolved SFH studies of additional interacting dwarfs are needed to determine observationally whether dwarf-dwarf interactions are a necessary or sufficient condition for flat (rather than outside-in) radial age gradients.

\subsection{Comparison to Simulations \label{compsimsect}}

\begin{figure}
\gridline{\fig{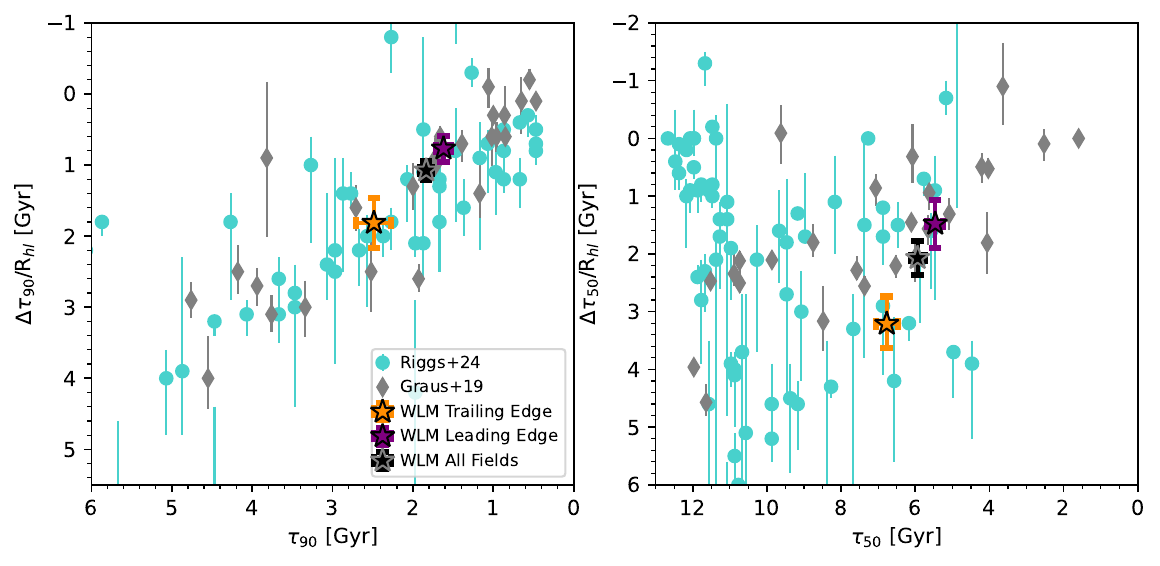}{0.99\textwidth}{}}
\caption{The radial gradient slope for $\tau_{90}$ (left) and $\tau_{50}$ (right) for our leading and trailing edge sample in WLM, compared to predictions from simulations by \citet{graus19} and \citet{riggs24}, shown as grey diamonds and cyan circles respectively.  Radial age gradient slopes are plotted versus the galaxy-wide values of $\tau_{90}$ or $\tau_{50}$ on the horizontal axis, and normalized to the half-light radius of each galaxy (see text for details).  Single values fit across all available WLM fields are shown in black.}  
\label{sim_fig}
\end{figure}

We compare our radial age gradients for WLM with predictions from two independent sets of cosmological zoom-in simulations by \citet{graus19} and \citet{riggs24}, discussed in Sect.~\ref{introgradsect}, in Fig.~\ref{sim_fig}.  The gradients we measure for both $\tau_{90}$ and $\tau_{50}$ are in good agreement with the simulations for both the leading and trailing edge sample, with the gradients measured from all of our available data lying predictably intermediate between the two.  
Clearly, results for a single galaxy cannot discriminate between the predictions of the two simulations, including the tighter correlation between global $\tau_{50}$ and its radial gradient from \citet{graus19} compared to \citet{riggs24}.  Nevertheless, this is the first time that radial gradients in $\tau_{90}$ and $\tau_{50}$ have been calculated for an isolated dwarf galaxy and compared to simulations, revealing excellent agreement.  However, a few caveats accompany these measurements.  

The comparison to simulations in Fig.~\ref{sim_fig} is not strictly direct for two reasons.  First, both sets of simulations measure radial gradients out to only 1.5R$_{hl}$, partially a practical restriction imposed by the stellar mass resolution of the simulations due to the need for sufficient particles to construct CSFHs in radial bins.  If we impose the same restriction, effectively excluding the NIRISS and UVIS fields, our results are unaffected to within their statistical uncertainties.  Second, while the global 
$\tau_{90}$ and $\tau_{50}$ values on the horizontal axis were calculated using all stars in the simulated galaxies\footnote{\textbf{This is not strictly true for the \citet{graus19} simulations.}  While both \citet{graus19} and \citet{riggs24} measured radial gradients out to 1.5R$_{hl}$, the global $\tau_{90}$ and $\tau_{50}$ values reported by \citet{graus19} were calculated excluding stars beyond 10\% of the virial radius of their simulated dwarfs, \textbf{while \citet{riggs24} did not impose any such restriction when calculating global $\tau_{90}$ and $\tau_{50}$ values.}}, this was not possible observationally for WLM due to incomplete spatial coverage.  \citet{graus19} used their simulations to evaluate biases in $\tau_{90}$ and $\tau_{50}$ measured at different galactocentric radii, finding that measurements at R$_{hl}$ were the best approximation to the global $\tau_{90}$ and $\tau_{50}$.  Therefore, we calculated global $\tau_{90}$ and $\tau_{50}$ values and uncertainties for WLM by evaluating our linear fits at R$_{hl}$=4.63$\arcmin$.  Assuming a different $R_{hl}$ \citep[e.g.,][]{hunter06,mcconnachie12} 
had no impact on the qualitative agreement with simulations in Fig.~\ref{sim_fig} since the older (younger) age at larger (smaller) R$_{hl}$ shifts our results nearly parallel to the predicted correlations in Fig.~\ref{sim_fig} rather than orthogonally.  

As a final caveat, both our measurement of radial gradient slopes (Fig.~\ref{radgrad_fig}, Table \ref{coefftab}) and those from the simulated galaxies \citep{graus19,riggs24} implicitly assume that radial stellar population gradients are linear.  At present, this is a practical requirement based on the quality of available observational data and the mass resolution of the simulations.  However, there are cases where radial age gradients appear non-linear, even when they are outside-in overall (e.g., Leo I; \citealt{ruizlara21}).  Observational constraints on the linearity of radial age gradients, only heretofore possible using hundreds of HST orbits \citep[e.g.,][]{williams09,cohen24} may be provided for the nearest dwarfs 
by the next generation of wide-field space-based imagers including the Nancy Grace Roman Space Telescope and Euclid \citep[e.g.,][]{hunt24}, complemented by JWST for more distant targets and/or earlier lookback times.

\section{Summary \label{summarysect}}

We have combined the NIRCam and NIRISS imaging fields towards WLM obtained by the JWST Resolved Stellar Populations ERS program with archival imaging of six additional HST fields (Fig.~\ref{obs_fig}, Table \ref{obstab}), all processed homogeneously to obtain PSF photometry catalogs (Sect.~\ref{photsect}, Figs.~\ref{jwstcmd_fig}$-$\ref{allcmd_fig}).  We divided each field into equally-populated radial bins, fitting a lifetime SFH to the CMD of each bin independently (Sect.~\ref{sfhfitsect}).  We measured radial age gradients by fitting per-bin values of $\tau_{90}$ and $\tau_{50}$ as a function of their projected semi-major axis equivalent distance R$_{SMA}$ (Sect.~\ref{resultsect}).  As a result, for the first time, we have:
\begin{itemize}
    \item Quantified the outside-in stellar age gradient in an isolated dIrr (Table \ref{coefftab}).  The age gradients we find in WLM are ($\delta$$\tau_{90}$, $\delta$$\tau_{50}$)/$\delta$R$_{SMA}=$(0.82$^{+0.10}_{-0.10}$, 1.60$^{+0.23}_{-0.22}$) Gyr/kpc assuming PARSEC evolutionary models.  
    \item Compared these measured radial gradients to predicted  $\tau_{90}$ and $\tau_{50}$ gradients in dwarfs from the latest cosmological simulations (Fig.~\ref{sim_fig}).  Although our observational result for one galaxy cannot discriminate between predictions from different simulations, we have provided the first direct validation of any internal radial stellar age gradients predicted by these simulations, showing good agreement.
    \item Measured a position-angle-dependent change in radial age gradients in WLM (Fig.~\ref{radgrad_fig}).  Fields towards the leading edge of WLM (i.e., in its direction of motion) have both younger \textit{values} of $\tau_{50}$ and $\tau_{90}$ and flatter \textit{radial gradients} of $\tau_{50}$ and $\tau_{90}$ versus R$_{SMA}$.
\end{itemize}

If ram pressure stripping by the IGM is confirmed as a means of triggering star formation in isolated dwarfs, there are two important implications.  First, the position-angle-dependent radial gradients we observe place a constraint that must be met by the latest simulations (e.g., those discussed in Sect.~\ref{compsimsect}).  Second, together with similarities in stellar kinematics \citep{leung21}, there is growing evidence that processes thought to be restricted to more dense environments (such as ram pressure stripping) are important for galaxies in isolation as well.  For the nearest Local Group galaxies, the Nancy Grace Roman Space Telescope may be needed to provide the requisite combination of contiguous spatial coverage and PSF stability to test this idea observationally.  However, more generally, radial age gradient measurements from statistically significant samples of dwarfs over a variety of environments are needed to test model predictions (e.g., Fig.~\ref{sim_fig}).  While radial age gradient measurements for a sample of 
nearby dwarfs with archival HST imaging 
will be presented elsewhere (R.~E.~Cohen, in prep.), it is clear (e.g., from Fig.~\ref{jwstcmd_fig}) that the efficiency of JWST for imaging studies beyond the Local Group is unmatched.  For example, assuming R$_{hl}$$\sim$1 kpc for dwarfs \citep[e.g.][]{mcconnachie12}, age gradients out to $\gtrsim$1.5R$_{hl}$ can be measured within a single JWST NIRCam module at D$_{\sun}$$\gtrsim$3 Mpc, well within the demonstrated capabilities of JWST for resolved stellar population studies \citep[e.g.,][]{newman24}.

\begin{acknowledgements}

It is a pleasure to thank Elena Sacchi for sharing published SFHs of NGC 4449 and UGC 4483 from \citet{sacchi18,sacchi21}, as well as the anonymous referee for their constructive, insightful comments.  This work is based in part on observations made with the NASA/ESA/CSA James Webb Space Telescope.  The data were obtained from the Mikulski Archive for Space Telescopes at the Space Telescope Science Institute, which is operated by the Association of Universities for Research in Astronomy, Inc., under NASA contract NAS 5-01327 for JWST.  These observations are associated with program ERS-1334.  This research is also based in part on observations made with the NASA/ESA Hubble Space Telescope obtained from the Space Telescope Science Institute, which is operated by the Association of Universities for Research in Astronomy, Inc., under NASA contract 5-26555.  These observations are associated with programs HST-GO-6798, HST-GO-6813, HST-GO-13768 and HST-GO-15275.  The JWST and HST data presented in this article were obtained from the Mikulski Archive for Space Telescopes (MAST) at the Space Telescope Science Institute.  The specific observations analyzed can be accessed via \dataset[doi: 10.17909/arvr-e123]{https://doi.org/10.17909/arvr-e123}. Support for this work was provided by NASA through grant ERS-1334 from the Space Telescope Science Institute, which is operated by AURA, Inc., under NASA contract 5-26555.  This research has made use of the NASA Astrophysics Data System Bibliographic Services.  R.~E.~C. acknowledges support from Rutgers the State University of New Jersey.  

\end{acknowledgements}

\facilities{JWST (NIRCam, NIRISS), HST (ACS/WFC, WFC3/UVIS, WFPC2)}
\software{astropy \citep{astropy}, matplotlib \citep{matplotlib}, numpy \citep{numpy}, emcee \citep{emcee}, DOLPHOT \citep{dolphot,dolphot2}, MATCH \citep{match}}

\end{document}